\documentclass{article}        
\usepackage{graphicx,graphics}           
\begin{document}

   \title{New statistical results on the optical IDV data of BL Lac  S5 0716+714}

   \author{B. D\u{a}nil\u{a}\footnote{Faculty of Physics, Babe\c{s}-Bolyai University, Cluj-Napoca, Romania}, A. Marcu\footnote{Faculty of Physics, Babe\c{s}-Bolyai University, Cluj-Napoca, Romania}, G. Mocanu\footnote{Faculty of Physics, Babe\c{s}-Bolyai University, Cluj-Napoca, Romania; gabriela.mocanu@ubbcluj.ro } \footnote{Department of Mathematics, Technical University of Cluj-Napoca, Romania}}

   \maketitle

\abstract{
This paper reports on the statistical behaviour of the optical IntraDay Variability of BL Lac S5 0716+714. Available IntraDay Variability data in the optical is tested to see whether or not the magnitude is log-normally distributed. It was consistently found that this is not the case. This is in agreement with a previous discussion for data for the same object but in a different observational period. Simultaneously, the spectral slope of the light curves is calculated. The implications of these findings for models which discuss both the location and the source of IntraDay Variability are presented.

\textbf{keywords}: radiation mechanisms: non-thermal, BL Lacertae objects: individual: S5 0716+714.}

\section{Introduction}           
\label{sect:intro}

BL Lac S5 0714+716 is an object with confirmed and documented variability in all wavelengths and on a wide interval of timescales. Theoretical tools have managed to satisfactorily explain most of the variability behaviour. However, there is a wide debate regarding the source of intra-day variability (IDV, variability in flux on a timescales of a few hours) detected in the light-curves (LCs) of such objects. The problem is at least two-fold: where is this source \textit{located}, with possible answers being the disc or the jet; and, more profoundly, what is the exact \textit{physical mechanism} able to produce such a fast variability. The answer to this final question is still incomplete, but there seems to be an agreement that the magnetic field is involved in some way. In fact, as shown in other astrophysical applications~(e.g., Lazarian \& Vishniac 1999, Oshuga et al. 2005) and somewhat successfully for these types of objects~(Mocanu et al. 2014), a model based on stochastic magnetic field reconnection provides many of the necessary properties that fit models to observations.

Bringing some light to both these issues must involve data analysis, modelling and reproducing (at least partially) the statistical properties exhibited by the light curves, e.g., the power spectral distribution (PSD) and the presence/absence of a linear flux versus root mean square relation (the rms-flux or, equivalently, the log normal distribution of the LCs)\footnote{From the point of view of their statistical behaviour, magnitude and flux are used interchangeably due to the fact that for the data considered here they are related through a linear relation.}. To our knowledge, this is the first simultaneous analysis of the spectral slope together with testing for a log-normal distribution hypothesis; we also analyse a set of simulated light curves with identical minimum, maximum, mean and variance as each of the observed LCs. These properties are discussed in connection to both the location and the nature of IDV in the next subsection; these general considerations are applied to observational data for BL Lac S5 0716+714 (Section~\ref{sect:analysis}) and the implications of the results are described in Section~\ref{sect:conclusion}.

\subsection{Location and nature of the IDV source}

There are two possible strong (theoretical) candidates for the location of the source of IDV: the disc (e.g., Hawley et al. 1996, Mineshige \& Yonehara 2001) or the jet (e.g., Chandra et al. 2011). Both theories have their advocates, and present both advantages and disadvantages. The advantage of both, from our point of view, is that in both situations a magnetic field is present. 

The continuum emission from accretion disks depends on the mass of the central black hole, such that low mass black hole produce continuum X-Ray emission and supermassive black holes (such as the object studied here) would produce a continuum in the optical (Czerny 2002, Frank et al. 2002). Optical IDV is seen as variations of output magnitude superimposed on the continuum emission (e.g., Gaskell \& Klimek 2003, Krolik 1999).

The data analysis in Section~\ref{sect:analysis} can help in at least two frameworks and is based on the following (explained in more detail in~Mocanu \& Bulcsu 2013)

\begin{enumerate}

\item{}historically, the optical/UV and X-Ray continua were thought to be partially connected through reprocessing in the disc~(Kawaguchi et al. 1998, Shakura \& Sunyaev 1973, Young et al. 2010)\footnote{However, it has been shown for particular AGN objects that the optical and X-Ray are not connected through reprocessing, not for long time scale variability, continua or IDV (McHardy et al. 2004, Gaskell \& Klimek 2003 and references therein).} and X-Ray variability does exhibit a linear rms-flux in its fast variability (Uttley et al. 2005, Gaskell \& Klimek 2003)\footnote{As an interesting side note, longer timescale optical variability (tens of days to years) does exhibit linear rms-flux relation (Gaskell \& Klimek 2003).}. Does the reprocessing change the distribution from log-normal to something else?

\item{}short time-scale variability in X-Rays (for low mass black holes) may be explained by the propagating fluctuations disc-model of~Lyubarskii (1997), which also naturally explains
both the PSD of the light curves and the linear rms-flux relation~(Arevalo et al. 2008, Scaringi et al. 2012). Conversely, it is believed that the existence of a linear rms-flux relation suggests that the variability originates in the accretion flow~(Arevalo \& Uttley 2006).

\end{enumerate}

We have recently proposed~(Mocanu \& Bulcsu 2013) that a valuable argument in this debate might be
offered by analysing whether or not optical IDV in AGNs shows a
linear rms-flux relation or, equivalently, to check if the LCs are log-normally distributed. For our previous set of data~(Mocanu \& Marcu 2012, Mocanu \& Bulcsu 2012), fast X-Ray variability for this BL Lac did exhibit a linear rms-flux, while optical IDV did not.

As observations show that the X-Ray and optical/UV flares are nonstationary and nonlinear, power spectra analysis alone does not adequately represent all the information contained in the light curve (Gaskell \& Klimek 2003, Uttley et al. 2005 gives a comprehensive discussion of Self Organized Criticality, PSD and the rms-flux relation for the X-Ray variability), so a joint magnitude distribution and PSD analysis is required.

The simplistic approach described above can fail and it is not conclusive if data analysis is not as extensive as possible. In the light of continuously developing models, it might be that IDV is produced in the disk and the process producing IDV cannot be fitted into the propagating fluctuations model. It it thus obvious that detailed analysis and discussion of statistical properties of the data is interesting in its own. Comparison of statistical data properties (e.g. PSD, linear/or-not rms-flux, flux distribution) coming from very different objects, like supermassive black holes, Solar mass black holes and the Sun itself show that light curves may share some statistical properties and have very different behaviours regarding others (Zhang 2007).

Analysis of data has shown that the source is stochastic~(Azarnia et al. 2005, Carini et al. 2011, Leung et al. 2011, Mocanu \& Marcu 2012) and it was shown that stochastic simulations can reproduce some of the characteristics of the data~(Harko \& Mocanu 2012, Mocanu et al. 2014, Xiong et al. 2000, Kawaguchi et al. 2000, Mineshige et al. 1994).

\section{Data analysis}
\label{sect:analysis}

\begin{figure}[h!]
   \centering
   \includegraphics[width=0.45\textwidth]{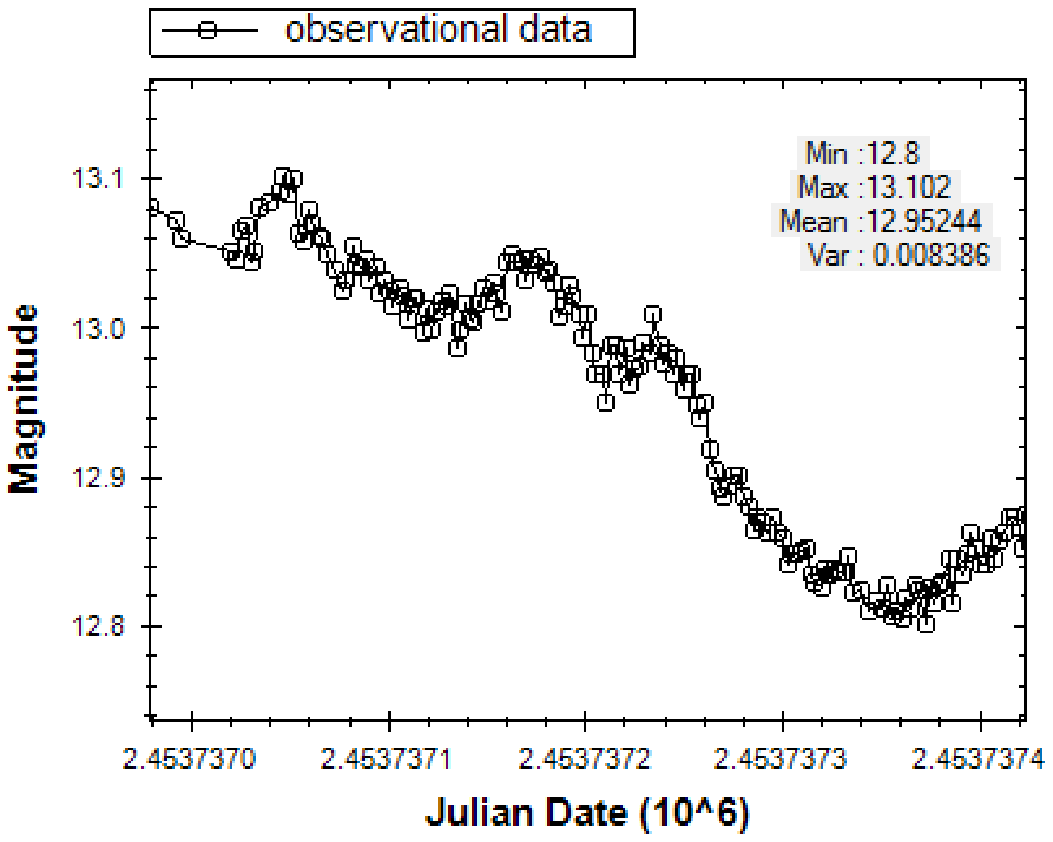}
   \includegraphics[width=0.45\textwidth]{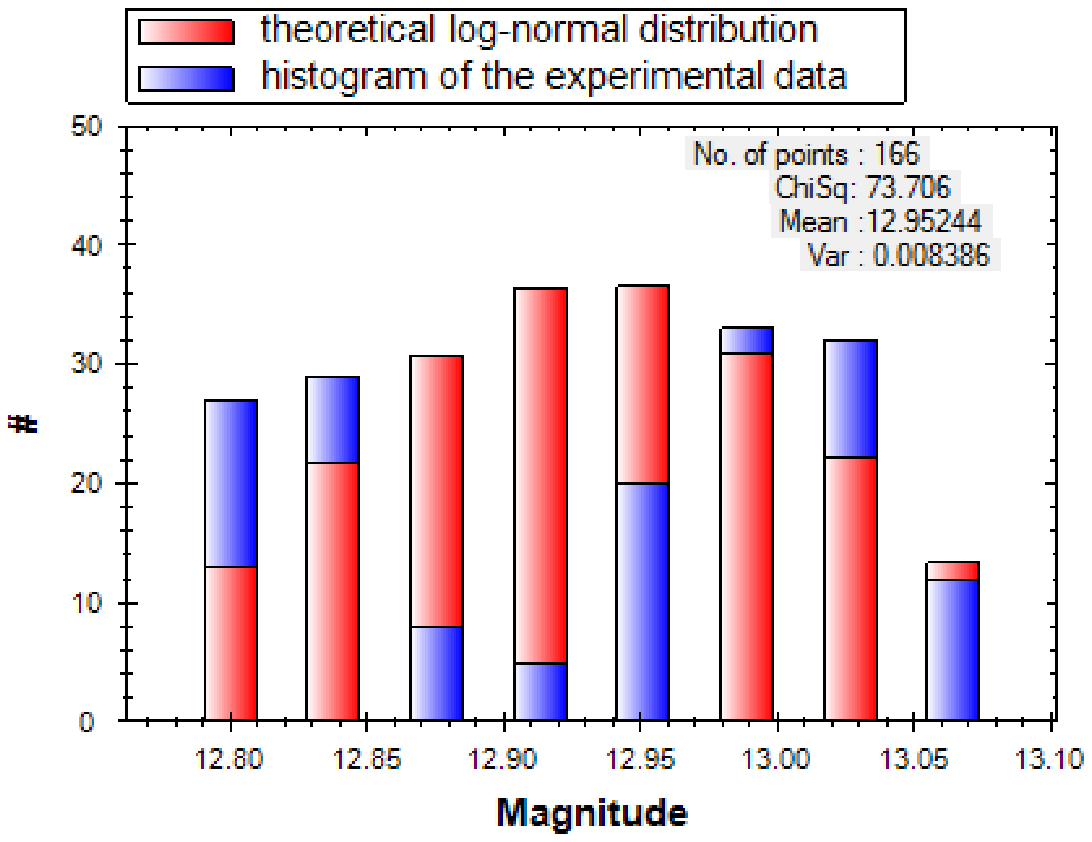}
   \caption{Left: Observed light curve. Right: Magnitude distribution of the light curve with the corresponding theoretical log-normal distribution superimposed.}
   \label{fig:obs1}
   \end{figure}

\begin{figure}[h!]
   \centering
   \includegraphics[width=0.45\textwidth]{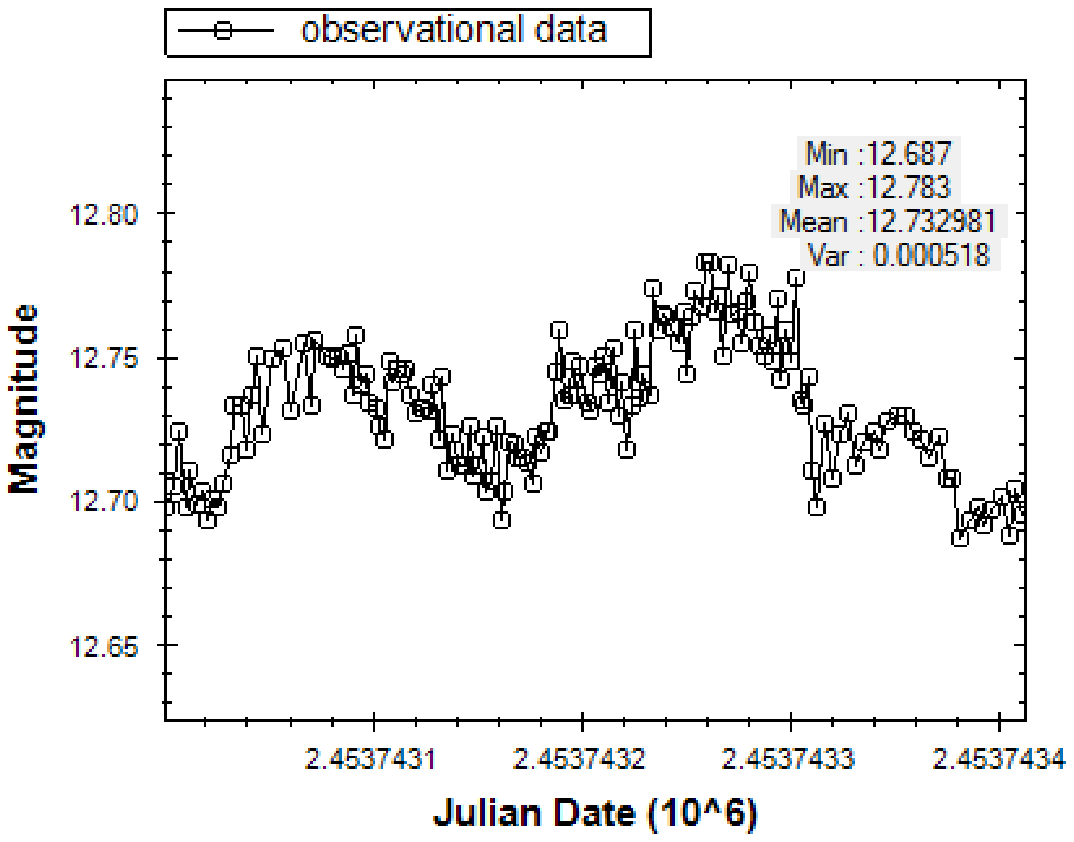}
   \includegraphics[width=0.45\textwidth]{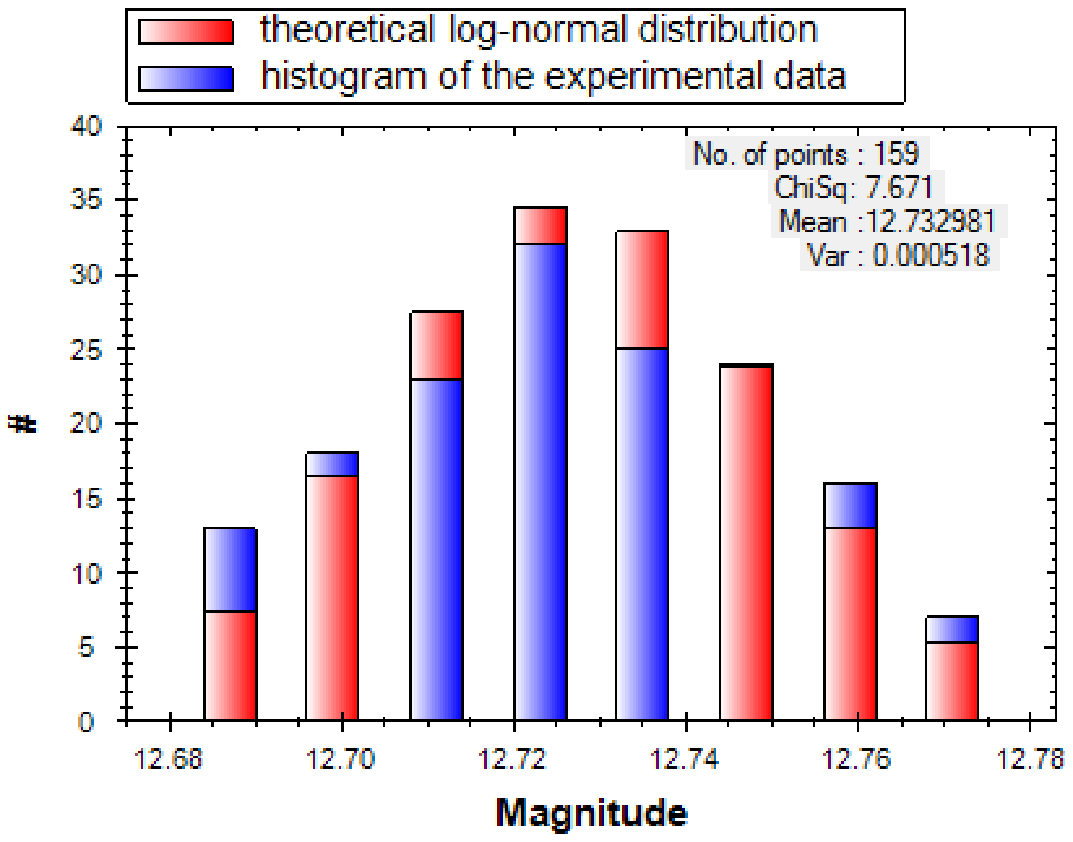}
   \caption{Left: Observed light curve. Right: Magnitude distribution of the light curve with the corresponding theoretical log-normal distribution superimposed.}
   \label{fig:obs2}
   \end{figure}

The data analysed has been previously discussed in~Wu et al. (2007) (examples of a light curves are shown in Figs.~\ref{fig:obs1} and \ref{fig:obs2} left). Observations have been carried out between 2006 January 1 and February 1. We discuss a set of 12 light curves which have at least 100 data points. The object was very active during this period, showing  variations with amplitudes larger than 0.1 mag and as large as 0.3 mag.

With a value of $z=0.31$ for the redshift of this object~(Nilsson et al. 2008) and a mass of $M=10^8M_\odot$~(Poon et al. 2009) for the central object, the extension of the emission region expressed in gravitational radii is $\Delta r=1.5\cdot\Delta t \cdot 10^{-3}$, where $\Delta t$ is the observed variability timescale. For $\Delta t = 1$ hour, the extension of the emission region is $5.5r_g$.

\begin{figure}[h!]
   \centering
   \includegraphics[width=0.45\textwidth]{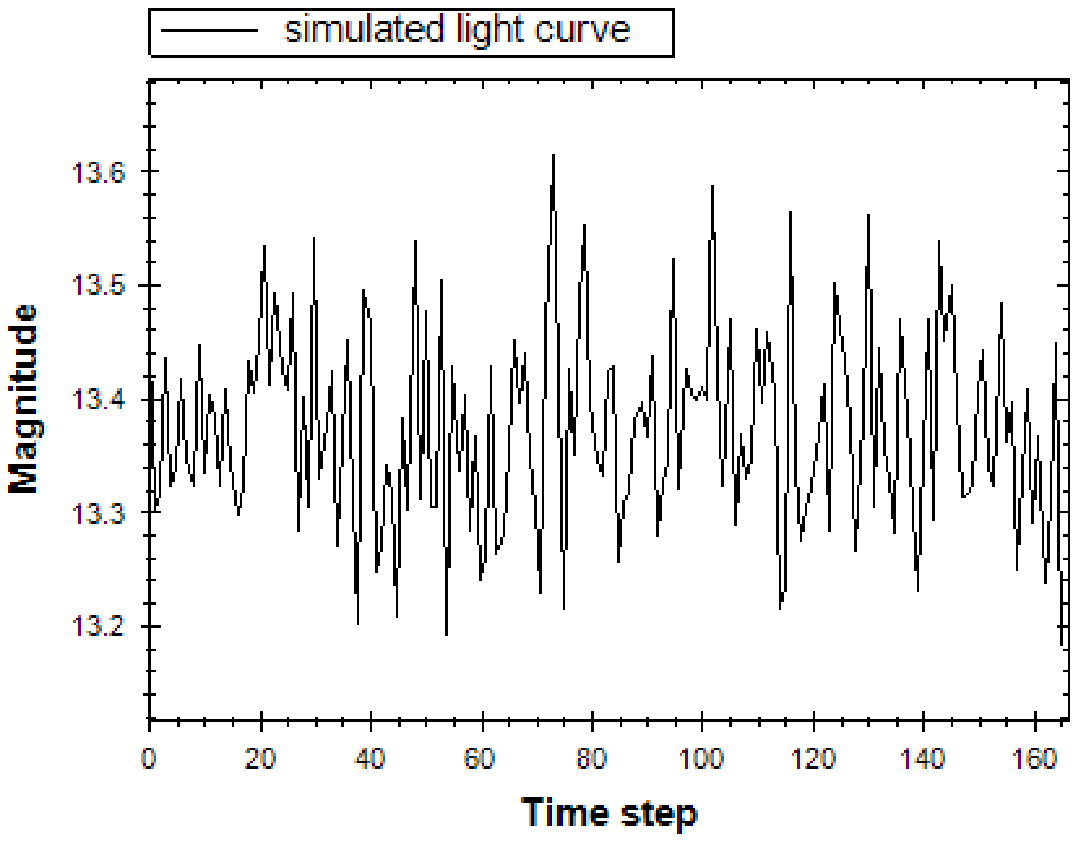}
   \includegraphics[width=0.45\textwidth]{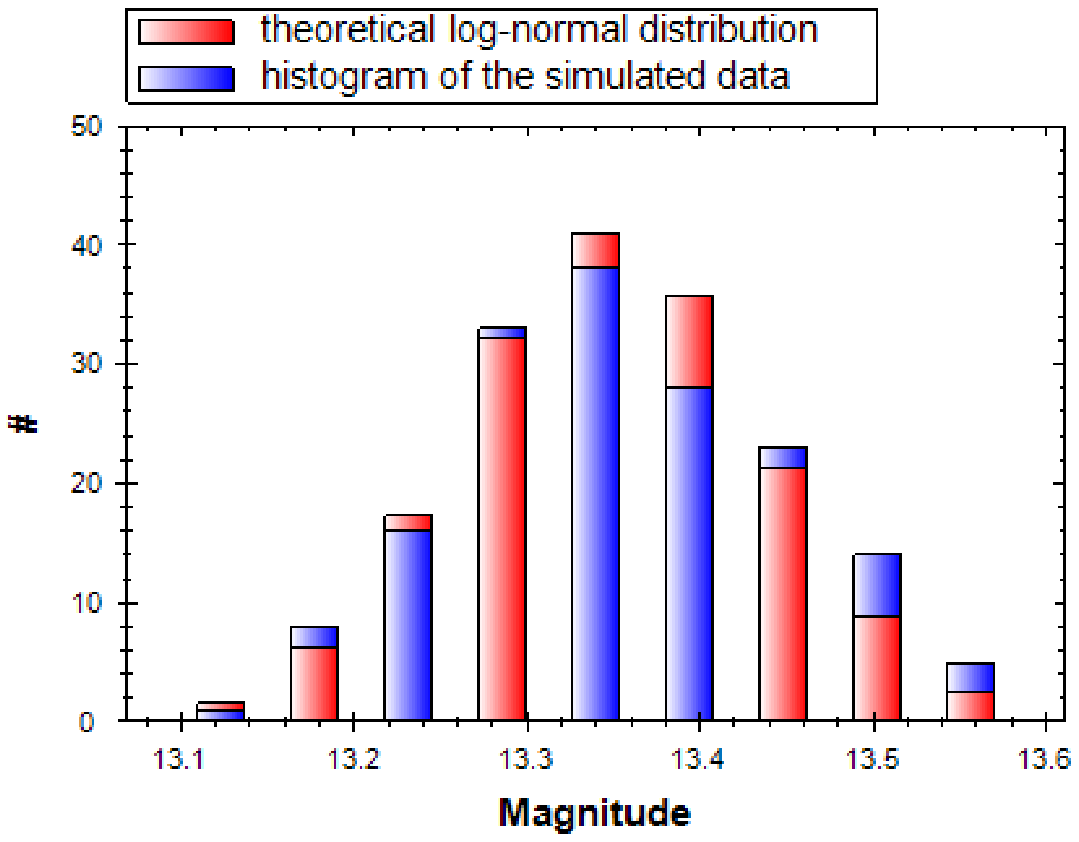}
   \caption{Left: Simulated light curve for a log-normally distributed process. Right: Magnitude distribution of this light curve with the corresponding theoretical log-normal distribution superimposed.}
   \label{fig:sim}
   \end{figure}

The procedure was previously used and described in~Mocanu \& Bulcsu (2013).  We tested if the light curves are log-normally distributed (e.g., Figs.~\ref{fig:obs1} and \ref{fig:obs2} right) by using the chi-square goodness of fit test. The constraint of having at least five members in each bin at all times was obeyed at all times, forcing us to use eight bins. The same $\chi ^2$ statistics was produced for a simulated light curve with a log normal distribution (Fig.~\ref{fig:sim}). The simulated light curve was built with constraints to have the same number of points as the observational data, as well as the same mean and variance.

The main results of the analysis are given in Table~\ref{tab:chi}. The first and second column provide the observation date and the band. The third column contains the $\chi ^2$ statistics for the experimental data, with the hypothesis that it is log-normally distributed and the fourth column shows the statistics for timeseries simulations. The fifth column contains the value of the spectral slope and the sixth the output Bayesian statistics value for the null hypothesis for the data and similarly the seventh and eighth columns for the simulated LCs. 

The spectral slope $\alpha$ and the Bayesian statistics $p_B$ are calculated with the .R software and the Bayes script described in detail in~Vaughan (2010). The value of the spectral slope $\alpha$ is calculated by mediating over many realizations of a process with the same characteristics as the light curve taken as input, realizations obtained through Monte Carlo simulations. The Bayesian probability assesses the correctness of the assumption, i.e. if $p_B$ is close to $1$, then the assumption that the emission process behaves so as to produce a luminosity with PSD $\sim f^{-\alpha}$ is correct. For details about the technical procedure, see~Vaughan (2010).

\begin{figure}[h!]
   \centering
   \includegraphics[width=0.45\textwidth]{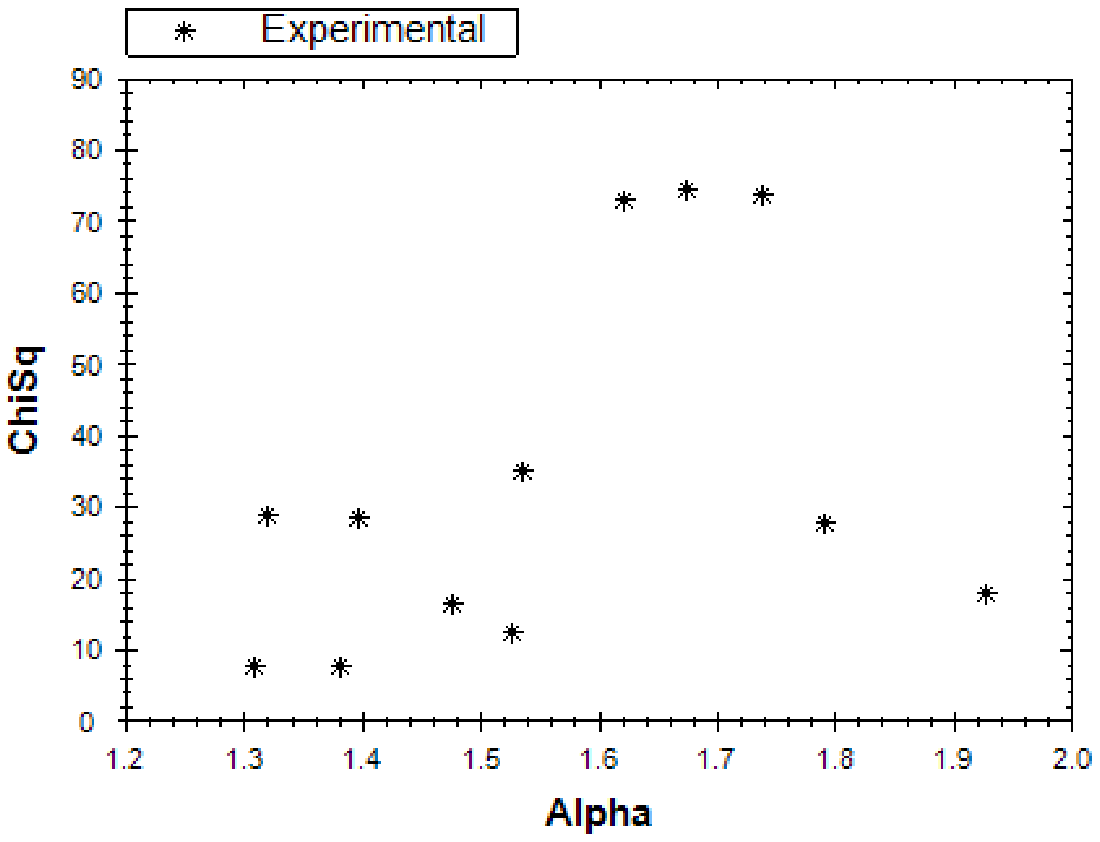}
   \includegraphics[width=0.45\textwidth]{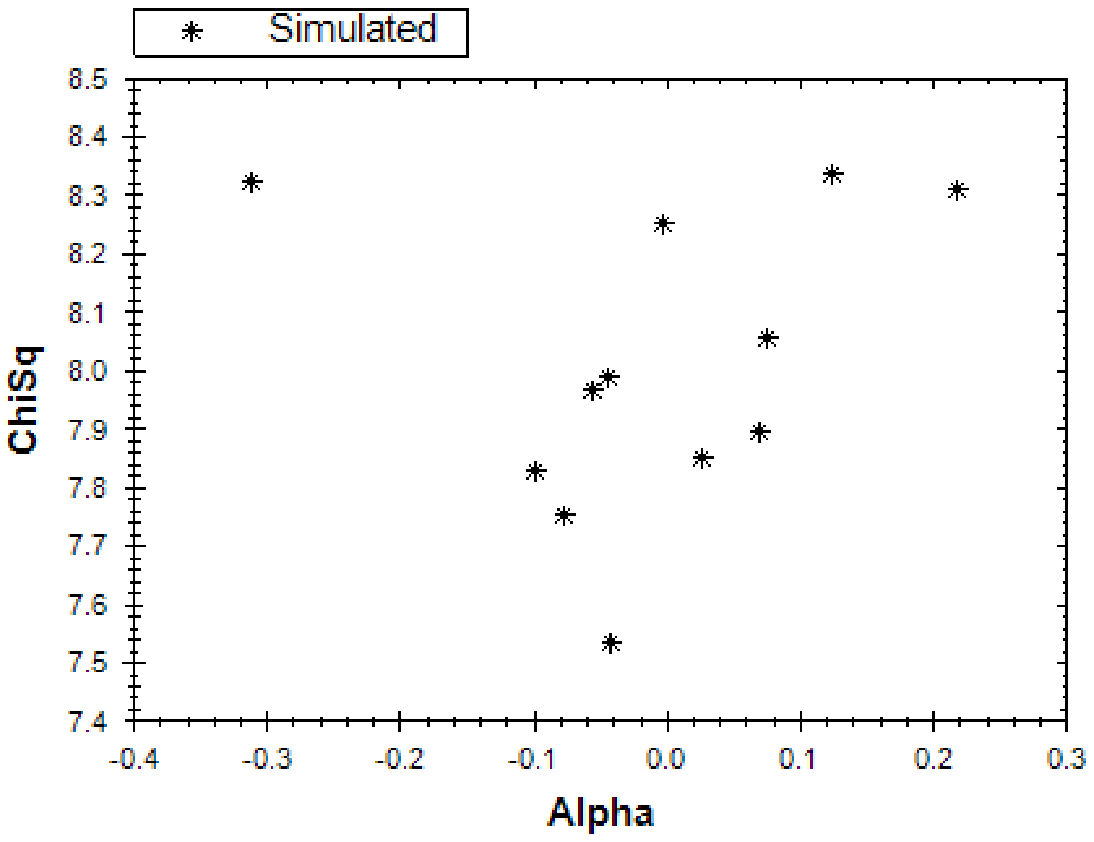}
   \caption{Spectral slope vs. calculated value for the $\chi ^2$ test for the observed light curves (left) and simulated light curves (right).}
   \label{fig:alphaVsChi}
   \end{figure}

The analysis of the results in the table clearly show that the data we analysed do not obey a log normal distribution. Even more, it can be seen that the needed values of $\alpha$ in order to produce an acceptable $\chi ^2$ are completely different from the observed ones (Fig.~\ref{fig:alphaVsChi}).

\section{Conclusions}
\label{sect:conclusion}

The spectral slope and the possibility that the optical IDV of Bl Lac S5 0716+714 is log-normally distributed was analysed. Based on this set of data and on a set analysed in a previous work~(Mocanu \& Marcu 2012, Mocanu \& Bulcsu 2012) we can conclude that it is more probable that the hypothesis of log normal distribution is false. A quick judge of this result might lead to the conclusion that this fact is an indication of IDV not being located in the disc. However, there is no proof that a log-normal distribution is a necessary and sufficient condition, but just eliminates some of the competing models as candidates for IDV. Although the discovered rms-flux in X-Ray cannot be obtained by standard shot noise models (although they do reproduce some PSD features) (Uttley \& McHardy 2001), these type of models are thus not excluded for optical fast variability.

Based on the wealth of data, interpretation, models and leaving room for the uncertainties inherent to all scientific endeavours pertaining to distant galaxies, this result tells us at least two things: from a numerical point of view, longer light curves are needed such that statistical analysis can reproduce conditions for the central limit theorem to hold; from a physics-based point of view, better models are needed such that the wealth of statistical characteristics of IDV light curves can be reproduced as accurately as possible within one single framework.

\subsection*{Acknowledgements}
This work was supported by a grant of the Romanian National Authority of Scientific Research, Program for research - Space Technology and Advanced Research - STAR, project number 72/29.11.2013.

\newpage

   \begin{table}[h!]
  \caption[]{ Results for the test of the hypothesis that the timeseries are log-normally distributed, with $\chi ^2 _{ref} = 11.07$, eight bins and five degrees of freedom. Results for the spectral slope under the hypothesis that the timeseries have the PSD $\sim f^{-\alpha}$. }
  \label{tab:chi}
  \begin{center}\begin{tabular}{cl|cl|cl|cl}
  \hline\noalign{\smallskip}
Date &  Band & $\chi ^2$(obs) & $\chi ^2$(sim) & $\alpha$(obs) & $p_B$(obs) & $\alpha$(sim) & $p_B$(sim)                  \\
  \hline\noalign{\smallskip}
2453736 & B & 28.589 & 8.251 & 1.398[0.104] & 0.539 & -0.002[0.117] & 0.57 \\
2453736 & V & 73.706 & 8.31 & 1.739[0.101] & 0.957 &  0.219[0.127] & 0.903\\
2453744 & V & 12.343 & 8.32 & 1.528[0.098] & 0.763 & -0.311[0.141] & 0.072\\
2453742 & R & 27.569 & 7.753 & 1.792[0.101] & 0.51 & -0.077[0.139] & 0.021\\
2453743 & R & 7.741 & 7.988 & 1.382[0.102] & 0.226 & -0.043[0.115] & 0.653\\
2453761 & R & 17.735 & 8.056 & 1.929[0.168] & 1 & 0.076[0.157] & 0.657\\
2453761 & V & 16.554 & 7.533 & 1.476[0.130] & 0.855 & -0.042[0.179] & 0.464\\
2453737 & V & 72.829 & 7.896 & 1.621[0.099] & 0.117 & 0.070[0.123] & 0.228\\
2453743 & V & 7.671 & 7.852 & 1.309[0.098] & 0.171 & 0.026[0.122] & 0.226\\
2453737 & R & 74.175 & 7.967 & 1.675[0.098] & 0.24 & -0.055[0.132] & 0.248\\
2453744 & R & 35.044 & 7.828 & 1.536[0.1] & 0.344 & -0.099[0.13] & 0.051\\
2453760 & V & 28.647 & 8.335 & 1.321[0.098] & 0.102 & 0.125[0.121] & 0.772                 \\
  \noalign{\smallskip}\hline
  \end{tabular}\end{center}
\end{table}

\end{document}